# Multispectral computational ghost imaging with multiplexed illumination

Jian Huang[A] · Dongfeng Shi*

Key Laboratory of Atmospheric Composition and Optical Radiation, Anhui Institute of Optics and Fine Mechanics, Chinese Academy of Sciences, Hefei 230031, China

Email:[A]Email:jhuang@aiofm.ac.cn,*dfshi@aiofm.ac.cn

**Abstract**

Computational ghost imaging is a robust and compact system that has drawn wide attentions over the last two decades. Multispectral imaging possesses spatial and spectral resolving abilities, is very useful for surveying scenes and extracting detailed information. Existing multispectral imagers mostly utilize narrow band filters or dispersive optical devices to separate lights of different wavelengths, and then use multiple bucket detectors or an array detector to record them separately. Here, we propose a novel multispectral ghost imaging method that uses one single bucket detector with multiplexed illumination to produce colored image. The multiplexed illumination patterns are produced by three binary encoded matrices (corresponding to red, green, blue colored information, respectively) and random patterns. The results of simulation and experiment have verified that our method can be effective to recover the colored object. Our method has two major advantages: one is that the binary encoded matrices as cipher keys can protect the security of private contents; the other is that multispectral images are produced simultaneously by one single-pixel detector, which significantly reduces the amount of the data acquisition.

**Keywords:** computational Imaging, multispectral and hyperspectral imaging, image reconstruction techniques

**1. Introduction**

Ghost imaging [1, 2] relied on the use of two correlated light fields and two detectors to create an image; one detector with no spatial resolution (such as photomultiplier tube) is used to collect the light field which has previously interacted with an object, and the other detector with high spatial resolution is employed to collect the other correlated light which never interacts with the object. Neither of the detectors alone can produce an image of the object, however combining the measurements made by both detectors can shape an image. Computational ghost imaging [3] is developed from ghost imaging. In computational ghost imaging system, the beam splitter and high spatial resolution detector are replaced by using a spatial light modulator capable of generating programmable light field to illuminate the scene. The intensity structures are calculated and stored in computer memory rather than measured by the detector with high spatial resolution. So only a single bucket photodetector is needed as imaging device in computational ghost imaging system. Single bucket detector has some significant virtues, such as high sensitivity, wide spectrum range, low cost, small size and light weight. So these virtues make computational ghost imaging has great potential in many fields. Recently, it has made great progresses and has already been put into practical applications, such as multilayer fluorescence imaging [4], optical encryption [5], remote sensing [6] and object tracking [7], etc.

Multispectral imaging which possesses spatial and spectral resolving abilities has drawn great attentions. L. H. Bian *et al* [8] proposed that utilizing the fast response of the detector, the 3D spatial-spectral information of the scene can be multiplexed into a dense 1D measurement sequence and then demultiplexed computationally under the single-pixel imaging scheme. Y. W. Wang *et al* [9] proposed a temporal multiplexing scheme for hyperspectral computational ghost imaging. The letter proposed a spectrum encoded acquisition scheme to achieve computational ghost imaging of hyperspectral data. Taking advantage of the speed gap between the extremely fast response of the bucket detector and magnitudes lower spatial illumination modulation, their approach temporally multiplexes a group of diverse spectra into the elapse of each 2-D illumination pattern. S. S. Welsh *et al* [10, 11] employed digital light projector in a computational ghost imaging system with multiple spectrally filtered photodetectors to simultaneously obtain the red, green and blue colored planes. The imaging system can obtain multi-wavelength reconstructions of real objects.

Most of the natural scenes have sparse characteristics under some transformations based on compression theory, and they can be recovered exactly from a relative small number of measurements. Furthermore, full sampling information of imaging object can be reconstructed under random sampling by employing compressed sensing algorithm [12]. In this Letter, we demonstrate a novel multispectral ghost imaging method that uses one single bucket detector with multiplexed illumination to produce colored image. The multiplexed illumination patterns are produced by three binary encoded matrices (corresponding to red, green, blue colored information, respectively) and random patterns. We investigated the proposed method numerically and experimentally. The results of simulation and experiment confirm our method is effective.

## 2. Method of multispectral image reconstruction

In general, iterative and compressed sensing algorithms are the two main types of reconstruction algorithms that can be employed in processing the acquired data in computational ghost imaging system. Iterative algorithm utilizes an entire data set in a bulk process to find the best solution for a set of unknowns [10]. While compressed sensing algorithms utilize less than 30% of Nyquist limit measurements to extract the objects [13]. So, here, we apply compressed sensing algorithm for our imaging reconstruction.

In our proposed method, the multiplexed illumination patterns are colored mixed. Firstly, three N×N encoded matrices named $E_R$, $E_G$ and $E_B$ are produced by computer program. And the three encoded matrices should abide by some properties as follows:
1). $E_R$, $E_G$ and $E_B$ are binary matrices.
2). $E_R + E_G + E_B = E$.
3). $E_m \cdot E_n = \begin{cases} 0 & m \neq n \\ E_m & m = n \end{cases} \quad m,n = R,G,B.$

That means the three encoded matrices are orthogonal and all the elements of E are equal to 1. For *i*-th projection, a random N×N pattern $S_i(x, y)$ is used to produce multiplexed illumination pattern $I_i$ which fuses the three products of the random pattern and the encoded matrices, is shown as follows:

$$I_i = E_R \cdot S_i(x,y) + E_G \cdot S_i(x,y) + E_B \cdot S_i(x,y). \tag{1}$$

$E_R\ S_i(x, y)$, $E_G\ S_i(x, y)$ and $E_B\ S_i(x, y)$ will be simultaneously loaded into the projection system. Then the multiplexed illumination pattern will be projected onto the object under certain frequency and the corresponding reflected intensity is detected by a single-pixel photodetector. The measured signal $U_i$ can be expressed as follows:

$$U_i = \sum_{N \times N} (C_R \cdot E_R \cdot S_i \cdot T_R + C_G \cdot E_G \cdot S_i \cdot T_G + C_B \cdot E_B \cdot S_i \cdot T_B). \tag{2}$$

Here, $T_R$, $T_G$ and $T_B$ are the red, green and blue information of the imaging scene, respectively. $C_R$, $C_G$ and $C_B$ represent correction coefficients of the single-pixel detector responding to the red, green and blue light, respectively, which can be gained by spectrum calibration. And the symbol $\sum_{N \times N}$ represents that the N×N elements of the matrix are summed. Eq. (2) can be rearranged as below:

$$U_i = \sum_{N \times N} S_i \cdot T. \tag{3}$$

Here, $T$ is the fused information of the imaging scene, and can be expressed as $T = C_R \cdot E_R \cdot T_R + C_G \cdot E_G \cdot T_G + C_B \cdot E_B \cdot T_B$. For M projections, Eq. (3) can be simplified as below:

$$S \cdot T = U. \tag{4}$$

Here, $U$ is a measurement signal vector of M×1; the random patterns $S$ are reshaped to M×N², and the object information $T$ is a vector of N²×1. In condition of M<<N, it seems hopeless to solve the object information $T$ since the number of equations is much smaller than the number of unknown variables. However, in most time, $T$ is compressible, and can be accurately recovered under condition of M<<N. Here, we use a second-order cone program of min-TV with quadratic constraints (available at *www.l1-magic.org*) to solve $T$. The constrained linear equation turns into the following optimization:

$$\min\ TV(T) \quad \text{subject to}\ \|S \cdot T - U\|_2 \leq \gamma. \tag{5}$$

The specified parameter $\gamma$ is set to 0.01. Once $T$ was to be got, $T_R$, $T_G$ and $T_B$ can be extracted by making use of the property of $E_R$, $E_G$ and $E_B$ as below:

$$\begin{aligned}
E_R \cdot T &= E_R \cdot (C_R \cdot E_R \cdot T_R + C_G \cdot E_G \cdot T_G + C_B \cdot E_B \cdot T_B) \\
E_G \cdot T &= E_G \cdot (C_R \cdot E_R \cdot T_R + C_G \cdot E_G \cdot T_G + C_B \cdot E_B \cdot T_B) \\
E_B \cdot T &= E_B \cdot (C_R \cdot E_R \cdot T_R + C_G \cdot E_G \cdot T_G + C_B \cdot E_B \cdot T_B)
\end{aligned} \tag{6}$$

Then, Eq. (6) can be sorted out and simplified as follows:

$$\begin{aligned}
Y_R &= E_R \cdot T = (C_R \cdot E_R) \cdot T_R \\
Y_G &= E_G \cdot T = (C_G \cdot E_G) \cdot T_G \\
Y_B &= E_B \cdot T = (C_B \cdot E_B) \cdot T_B
\end{aligned} \tag{7}$$

Here, $Y_R$, $Y_G$ and $Y_B$ are known N×N matrices which can be reshaped to N²×1 matrices, respectively. And $C_R\ E_R$, $C_G\ E_G$ and $C_B\ E_B$ are also known N×N matrices which can be reshaped to N²×N² diagonal matrix. $T_R$, $T_G$ and $T_B$ are N²×1 matrices. In order to exactly capture the full sampling information of imaging object, optimizations can be resorted as follows:

$$\min\ TV(T_m) \quad \text{subject to}\ \|(C_m \cdot E_m) \cdot T_m - Y_m\|_2 \leq \delta \quad m = R, G, B. \tag{8}$$

Here, the specified parameter $\delta$ is set to 0.01. At last, $T_R$, $T_G$ and $T_B$ are reshaped to N×N matrices, and the final reconstructed colored image can be produced by fusing $T_R$, $T_G$ and $T_B$.

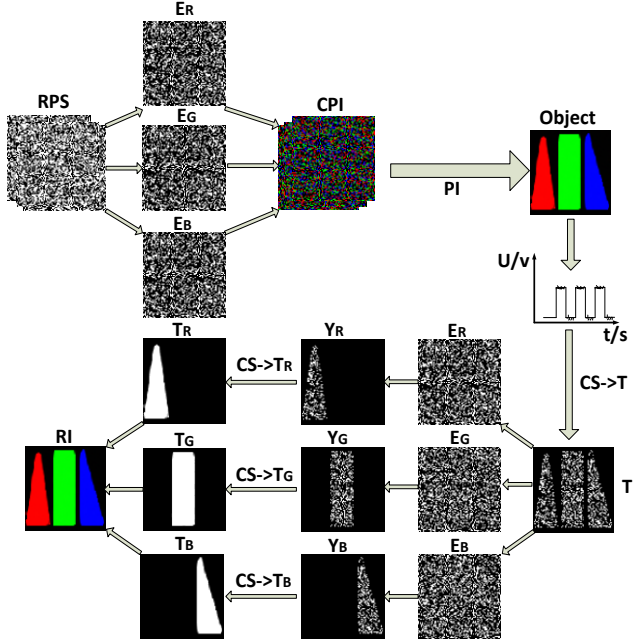

**Figure 1**. Procedure of our method for reconstructing object. RPS: random patterns, *S*; CPI: colored patterns, *I*; PI: projection illumination; CS->T: solve T based on Eq. (5); CS->$T_R$, CS->$T_G$ and CS->$T_B$: compute $T_R$, $T_G$, and $T_B$, based on Eq. (8); RI: recovered image.

The whole procedure of the proposed method is demonstrated in Figure 1. First, computer program produces three binary encoded matrices $E_R$, $E_G$, $E_B$ and random patterns (RPS) S. The multiplexed illumination colored patterns (CPI) I are produced by fusing the product of random patterns and the three encoded matrices. They are loaded into the projection system, and then projected onto the scene. The reflected light intensity U is collected by a single-pixel photodetector. The fused information of the scene T can be restored by Eq. (5). Then, T is separated into three parts which are named $Y_R$, $Y_G$ and $Y_B$, indicating red, green and blue information under partial random sampling, respectively. The full sampling information of red, green and blue reflected of the scene named $T_R$, $T_G$ and $T_B$ can be got by Eq. (8). At last, the colored image will be reconstructed by integrating the full sampling information of red, green and blue reflected information.

## 3. Experimental verification

3.1 Quantitative Research on Simulation

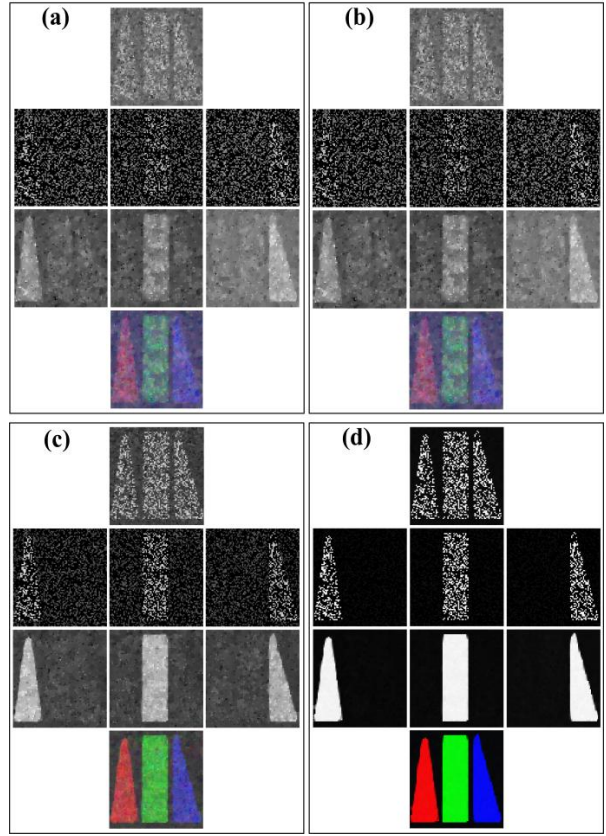

**Figure 2**. Simulation results of imaging to a colored object under different measurements. (a) The result of 3000 measurements. (b) The result of 3500 measurements. (c) The result of 4000 measurements. (d) The result of 4500 measurements.

In order to evaluate our method, we design a colored object (the pixel resolution is 81×81) containing three parts: a red isosceles triangle, a green rectangle and a blue right triangle. Simulation results are shown in Figure 2 which demonstrates recovered results under 3000, 3500, 4000 and 4500 measurements, respectively. The first row is the fused information of the imaging object $T$. The results of red, green and blue information under partial random sampling in gray level are shown in the second row (from left to right), which are separated by the three encoded matrices $E_R$, $E_G$, $E_B$ and the information $T$, respectively. The third row (from left to right) is the full sampling results of red, green and blue information of colored object in gray level. The fourth row is the final reconstruction result of the imaging object. In this letter, the number of 1 in the three binary matrices $E_R$, $E_G$ and $E_B$ are equal to 2187 which is one-third of total pixels of imaging object. Of course, the number of 1 in the three binary matrices needn't be equal in some other situation. From the simulation results, we can see the recovered quality get better when illumination patterns increase. Figure 3 is the plot of RMSE (root mean square error) with respect to pattern numbers for quantitative analysis for our method. The RMSE of $T_R$, $T_G$ and $T_B$ are almost constant while measurements exceed 5000. The results of simulation show that our method can distinguish the specific colored information and reconstruct the colored object via multiplexed illumination.

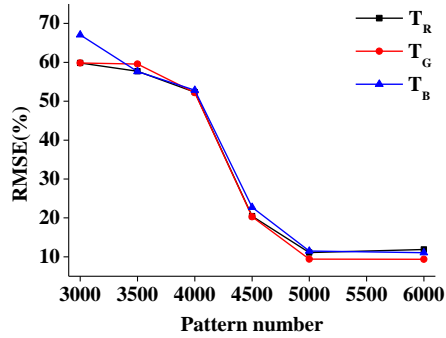

**Figure 3.** The plot of RMSE with respect to pattern numbers.

3.2 Experimental Study

Next, an experimental system is built in the laboratory environment to test our method .The imaging object is depicted in Figure 4 (a). The red, green and blue encoded matrices are shown in Figure 4(b), Figure 4(c) and Figure 4(d), respectively. The multiplexed illumination patterns which were also employed in simulation experiment are projected to the imaging object. The multispectral computational ghost imaging setup is shown in Figure 5. The digital projector (SONY 3-LCD VPL-CX131) illuminated the object with the multiplexed patterns. The 3-LCD projector contains an ultrahigh voltage mercury lamp which produces white light; two dichroic mirrors can separate the white light into red, green and blue light which will transmit onto corresponding LCDs. The LCDs can be programmed by computer programmer, respectively. The detection unit (a photomultiplier, THORLABS PMM02-1, 280-850nm) collected the reflected light from the object. An analogue to digital converter (NI USB-6211, maximum sampling rate is 250k/s for single channel) is used to digitize the detected signal and a computer is employed to generate the illumination patterns and perform multispectral reconstructions of the test object. The corresponding experimental results are shown in Figure 6. The pixel resolution of imaging is 81×81.The rank sequences of the experimental results are as same as the simulation results in Figure 2. From the experimental results shown in Figure 6, we can find that the quality of the recovered results enhance along with measurements increase. The experimental results confirm that our method is effective. However, there are some differences between the recovered results of simulation and experiment. The reasons may be as follows：1). The imaging object is binary in simulation, while the experimental imaging object is gray; 2). Maybe there are some disturbances (such as electronic noise, stray light, etc.) in experimental environment, while the noises are neglected in simulation.

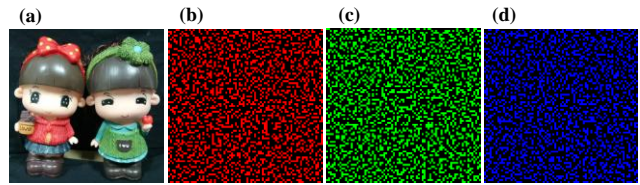

**Figure 4**. (a) The imaging target. (b) The encoded matrix of red showing in red level. (c) The encoded matrix of green showing in green level. (d) The encoded matrix of blue showing in blue level.

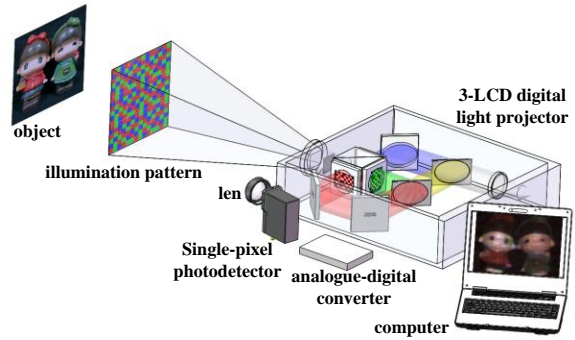

**Figure 5**. Schematic of our experimental system.

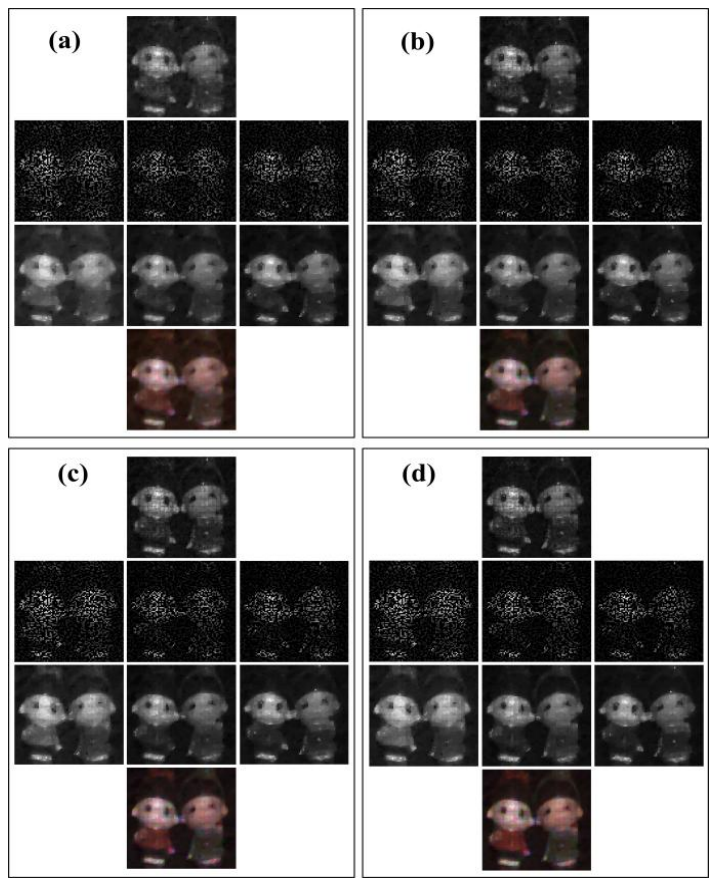

**Figure 6**. Experimental results of imaging to two adjacent colored kids under different measurements. (a) The result of 3000 measurements. (b) The result of 4000 measurements. (c) The result of 5000 measurements. (d) The result of 6000 measurements.

The results of the simulation and experiment have shown that our method can recover colored object and the quality of the recovered multispectral image get better when measurements increase. Of course, the quality of recovered image can be further improved by other approaches, such as application of 2D orthogonal sinusoidal patterns instead of random patterns in Ref. [14]. Our method has some superiority: one is that multispectral images are produced simultaneously by one single-pixel detector which makes our system more compact and robust; the other one is that the red, green and blue information are encoded by the three binary matrices which are very crucial for recovering multispectral image. The imaging object can't be exactly recovered if the encoded matrices are inaccurate. Therefore, our proposed method can protect the security of private contents, and could also be applied in information encryption.

## 4. Conclusions

In summary, we have proposed and validated a novel multispectral computational ghost imaging method with multiplexed illumination. The multiplexed illumination patterns are produced by three binary encoded matrices (corresponding to red, green and blue colored information, respectively) and random patterns. In order to exactly recover the multispectral imaging object, compressed sensing algorithms are employed four times in our reconstructed procedure; so there is still a little time consume. The next focus is to improve efficiency of the recovered algorithms. Besides, in the simulation and experiment, we have found that imaging quality is easily influenced by the three binary encoded matrices which are crucial to accomplish multiplexed illumination; so the other further study is to optimize the encoded matrices.

### Acknowledgement

This work was supported by the National Natural Science Foundation of China (No. 11404344, 41505019) and CAS Innovation Fund Project (No. CXJJ-17S029).